\documentclass[aps,prb,showpacs,twocolumn,floats]{revtex4}
\usepackage{graphicx}
\usepackage{subfigure}
\usepackage{epsfig}
\usepackage{dcolumn}
\usepackage{bm}
\usepackage[ansinew]{inputenc}
\usepackage{amsmath}
\usepackage{amsthm}
\usepackage[T1]{fontenc}
\usepackage{amssymb}
\usepackage{amsfonts}
\usepackage[english]{babel}
\usepackage{enumitem}

\begin{document}

\title{Resonant Spin Tunneling in Randomly Oriented Nanospheres of Mn$_{12}$ Acetate}
\date{\today}
\author{S. Lend\'{i}nez, R. Zarzuela, J. Tejada}
\affiliation{Departament de F\'{i}sica Fonamental, Facultat de F\'{i}sica, Universitat de Barcelona, Mart\'{i} i Franqu\`{e}s 1, 08028 Barcelona, Spain}
\author{M. W. Terban$^1$, S. J. L. Billinge$^{1,2}$}
\affiliation{$^1$Department of Applied Physics and Applied Mathematics, Columbia University, New York, NY 10027, USA\\$^2$Condensed Matter Physics and Materials Science Department, Brookhaven National Laboratory, Upton, NY 11973, USA}
\author{ J. Espin$^1$, I. Imaz$^1$, D. Maspoch$^{1,2}$}
\affiliation{$^1$Institut Catal\`{a} de Nanotecnologia, ICN2, Esfera Universitat Aut\'{o}noma Barcelona (UAB), Campus UAB, 08193 Bellaterra,  Spain\\$^2$Instituci\'{o} Catalana de Recerca i Estudis Avan\c{c}ats (ICREA), 08100 Barcelona, Spain}
\author{E. M. Chudnovsky}
\affiliation{Physics Department, Lehman College,
The City University of New York, 250 Bedford Park Boulevard West, Bronx, NY 10468-1589, USA}

\begin{abstract}
We report measurements and theoretical analysis of resonant spin tunneling in randomly oriented nanospheres of a molecular magnet. Amorphous nanospheres of Mn$_{12}$ acetate have been fabricated and characterized by chemical, infrared, TEM, X-ray, and magnetic methods. Magnetic measurements have revealed sharp tunneling peaks in the field derivative of the magnetization that occur at the typical resonant field values for Mn$_{12}$ acetate.  Theoretical analysis is provided that explains these observations. We argue that resonant spin tunneling in a molecular magnet can be established in a powder sample, without  the need for a single crystal and without aligning the easy magnetization axes of the molecules. This is confirmed by re-analyzing the old data on a powdered sample of non-oriented micron-size crystals of Mn$_{12}$ acetate. Our findings can greatly simplify the selection of candidates for quantum spin tunneling among newly synthesized molecular magnets. 
\end{abstract}


\maketitle

\section{Introduction}\label{introduction}

Spin tunneling in molecular magnets has been subject of intensive research in the last 20 years \cite{MQT-book,Springer}. These systems exhibit unique quantum features that show up in macroscopic experiments. Among them are stepwise quantum hysteresis due to resonant tunneling at discrete values of the magnetic field \cite{Friedman-PRL1996,Hernandez-EPL1996,Barbara-Nature1996}, topological Berry phase effects \cite{WS}, magnetic deflagration \cite{CCNY-PRL2005,UB-PRL2005}, and Rabi oscillations \cite{Schledel-PRL2008,Bertaina-Nature2008}. Most recently, experiments with individual magnetic molecules bridged between conducting leads and molecules grafted on carbon nanotubes have been performed that permit readout of quantum states of individual atomic nuclei \cite{Wern-NatureNano2013,Wern-ASC-Nano2013}. Quantum superposition of spin states in magnetic molecules makes them candidates for qubits -- elements of quantum computers \cite{Nanotech-2001}. 

Beginning with the discovery of resonant spin tunneling in Mn$_{12}$ acetate \cite{Friedman-PRL1996} it has been generally believed that at the macroscopic level the effect can only be observed in a single crystal or in the array of microcrystallites whose crystallographic axes are aligned. The latter method was used in Refs. \onlinecite{Friedman-PRL1996,Hernandez-EPL1996}, and in a few subsequent publications by the same authors, before sufficiently large single crystals of molecular magnets had become available. Nowadays a large number of new molecular magnets are synthesized by chemists every year and subsequently tested by physicists for the presence of spin tunneling. In many cases growing a sufficiently large single crystal is a challenging task as compared to growing microcrystals. At the same time the effectiveness of the room-temperature alignment of microscrystals by a high magnetic field in, e.g., an epoxy strongly depends on the shape of the crystallites and may also be incompatible with their chemistry. 

In this paper we show that neither macroscopic single crystals or the alignment of microcrystals are necessary to obtain solid evidence of spin tunneling in macroscopic magnetic measurements of molecular magnets. This can be done by simply measuring the magnetization curve, $M(H)$, and plotting the derivative of the magnetization on the magnetic field, d$M$/d$H$. To illustrate this point we use amorphous nanospheres of Mn$_{12}$ acetate \cite{Imaz-ChemCom2008,Carbonera-ICA208}. Such nanospheres neither form crystals nor they allow to align easy magnetization axes of Mn$_{12}$ molecules in a high magnetic field because of their isotropic magnetic susceptibility.  

First magnetic measurements of similar Mn$_{12}$ nanoparticles were performed in Refs. \onlinecite{Imaz-ChemCom2008,Carbonera-ICA208} where peaks in d$M$/d$H$ were observed and attributed to quantum tunneling. However, no explanation to such observation in a disordered sample has been provided. Zero-field cooled magnetization curve of the measured sample revealed large, up to $40$\%, fraction of fast-relaxing species of Mn$_{12}$ acetate. This resulted in a complicated pattern of displaced tunneling maxima as compared to typical resonances in a Mn$_{12}$ crystal that are separated by about $4.6$ kOe. 

Highly amorphous Mn$_{12}$ particles used in our experiments did not show any visible presence of the second species. Sharp tunneling maxima in d$M$/d$H$ have been observed. Their location coincided with the position of tunneling maxima in a Mn$_{12}$ crystal. We provide theoretical analysis that explains observation of tunneling resonances in a fully disordered sample. The predicted post-resonance field dependence of the peaks agrees with experiment. The practical value of this observation  is that it can greatly reduce the preliminary work by chemists that is required to search for quantum spin tunneling in newly synthesized molecular magnets. As long as the magnetic molecules preserve their structure, the disordered sample of crystallites of any size is sufficient for that task. 

The paper is structured as follows. Fabrication and characterization of the samples by various techniques are discussed in Section \ref{fabrication}. Measurements of the field and temperature dependence of the magnetization, and of the ac susceptibility, are presented in Section \ref{magnetic}. They are analyzed and explained in Section \ref{analysis}. Section \ref{discussion} summarizes our results and offers some final remarks. 

\section{Fabrication and characterization of Mn$_{12}$ nanospheres}\label{fabrication}

To obtain spherical Mn$_{12}$ acetate particles showing the highest degree of amorphous character they were synthesized using a method adapted from the earlier publication \cite{Imaz-ChemCom2008}. In a typical experiment, $60$ mg of Mn$_{12}$ acetate were dissolved in $15$ mL of acetonitrile. The solution was added to $30$ mL of toluene under vigorous stirring. Precipitation of a brown solid was observed. One hour later the resulting dispersion was centrifuged ($4' \times 8000$ rpm) and the supernatant was collected. The pellet was redispersed with acetonitrile:toluene ($1:2$), centrifuged again and the supernatant was collected again. Both supernatants were then mixed and dried under vacuum condition. $5.8$ mg of a brown solid ($10 \%$ yield) were collected and stored in vacuum. 

\begin{figure}[htbp!]
\includegraphics[width=14cm,angle=0]{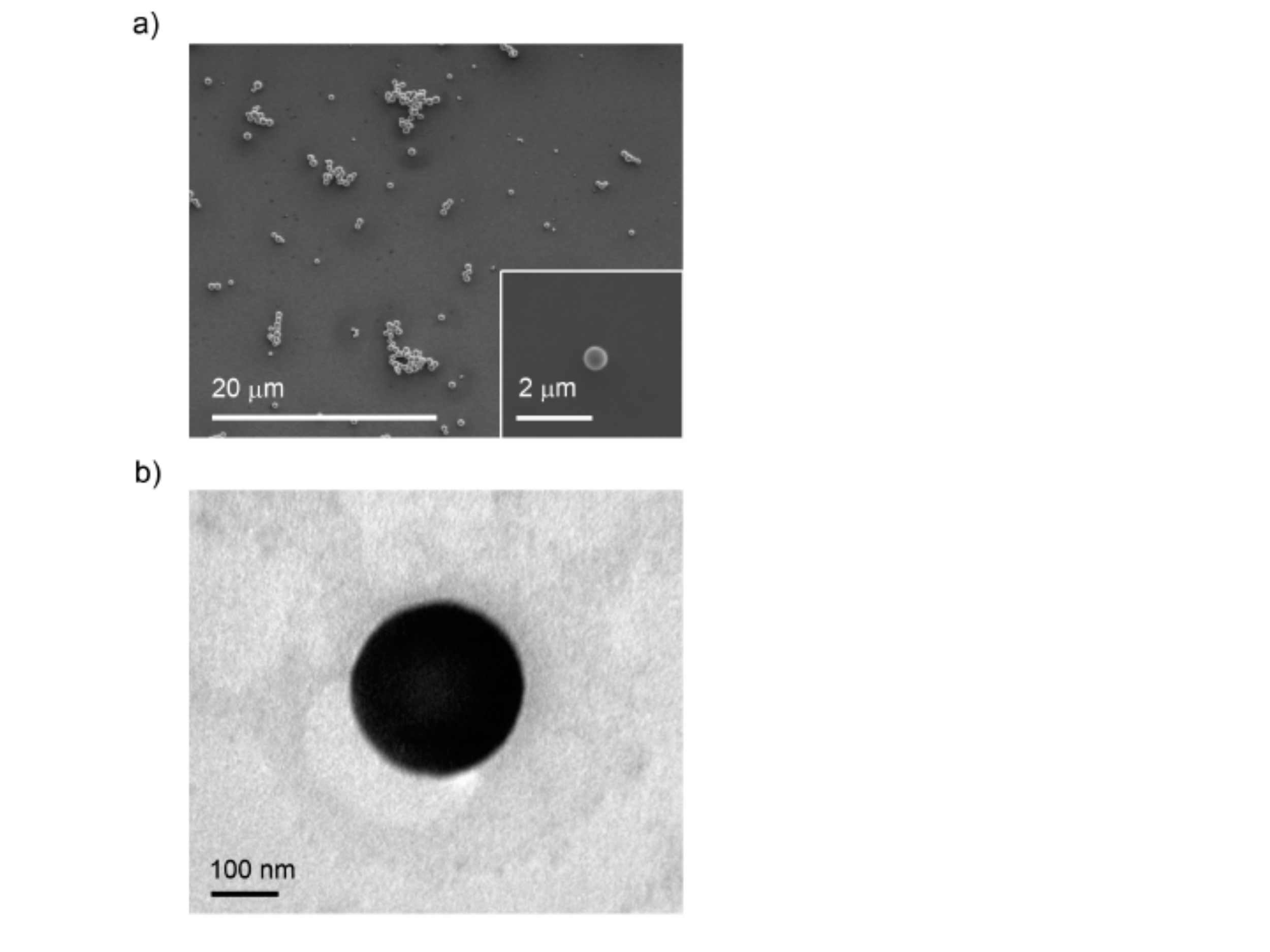}
\caption{ a) Representative SEM image of Mn$_{12}$ acetate spherical particles. Inset: individual particle. b) TEM image of an individual Mn$_{12}$ acetate nanosphere.}
\label{fig particles}
\end{figure}
\begin{figure}[htbp!]
\centerline{
\includegraphics[width=10cm,angle=0]{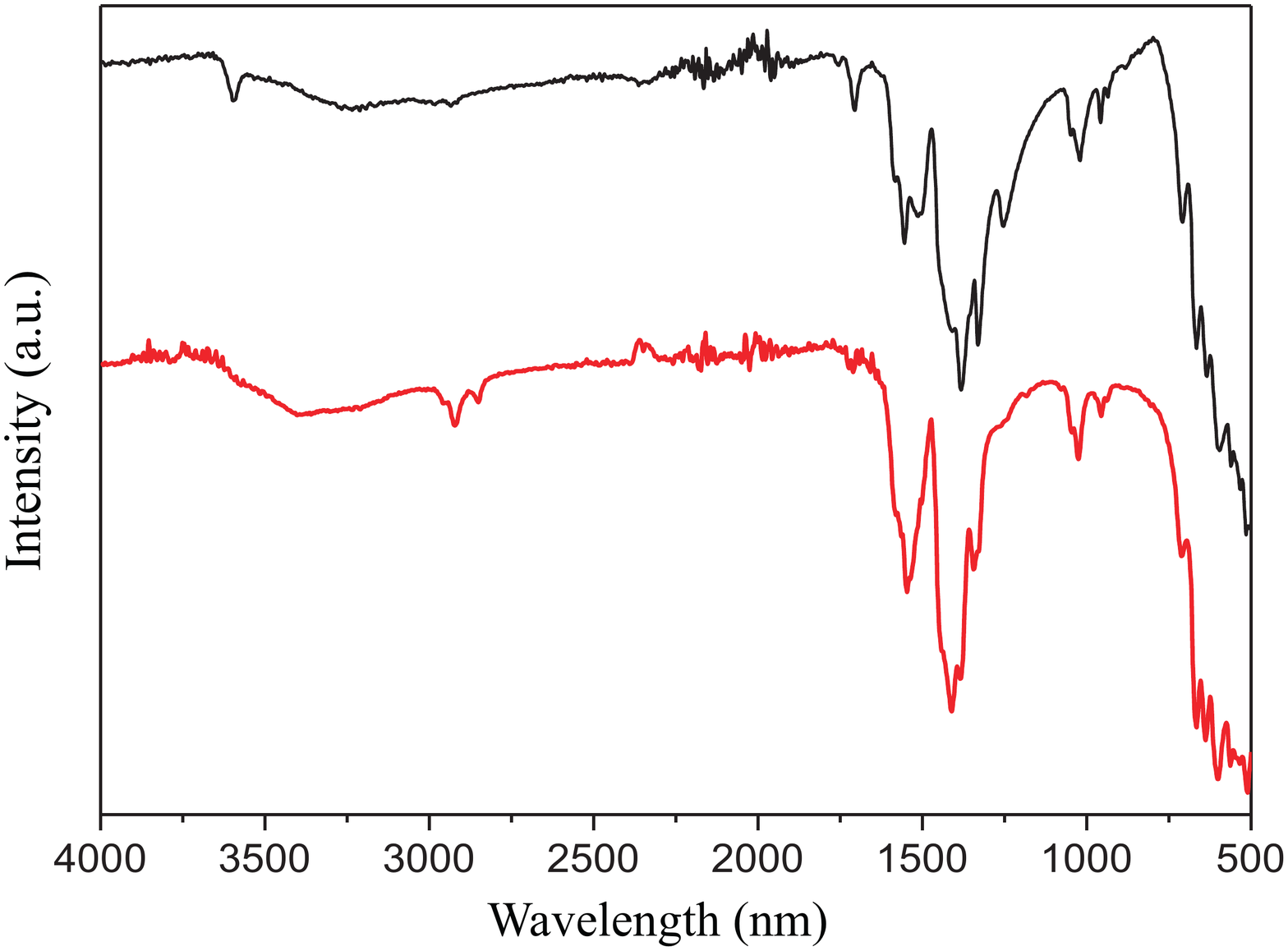}}
\caption{Color online: IR spectra of bulk Mn$_{12}$ acetate crystals (black) and amorphous Mn$_{12}$ acetate nanospheres (red). }
\vspace{1.5cm}
\label{fig IR}
\end{figure}
Scanning and Transmission Electron Microscopy (SEM and TEM) images of the brown solid revealed the formation of uniform spherical particles, see Fig. \ref{fig particles}. The size of the particles was calculated from FESEM images by averaging the diameter of at least $300$ particles from different areas of the sample. The average size of 237 $\pm$ 69 nm and the median size of $238$ nm were determined. This average size was further confirmed by dynamic light scattering (DLS). The chemical correspondence with bulk [Mn$_{12}$O$_{12}$(CH$_3$COO)$_{16}$(H$_2$O)$_4$] crystals was confirmed by elemental analysis and by the positive matching between the IR spectra, see Fig. \ref{fig IR}. As we shall see below, the additional evidence of the conventional structure of Mn$_{12}$ molecules in the nanospheres follows from the magnetic measurements that reveal the resonant spin tunneling at the same values of the magnetic field as for the single crystals of Mn$_{12}$ acetate. 

\begin{figure}[htbp!]
\includegraphics[width=8.5cm,angle=0]{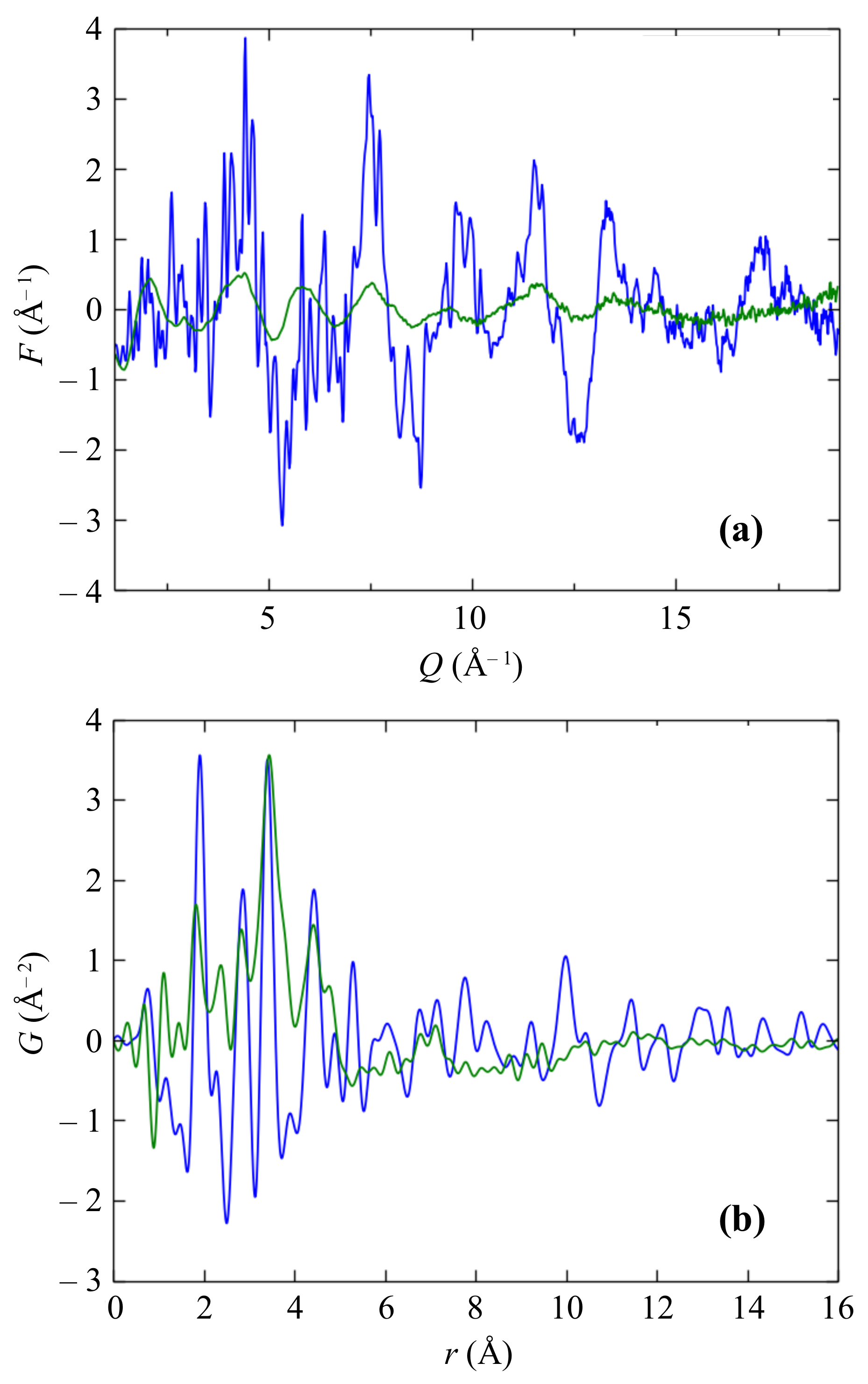}
\caption{Color online: (a) Structure factors for microscrystals (blue) and nanospheres (green) of Mn$_{12}$ acetate. (b) Superimposed pair distribution functions for microscrystals (blue) and nanospheres (green) of Mn$_{12}$ acetate.}
\label{fig structure}
\end{figure}
To confirm the amorphous character of spherical Mn$_{12}$ particles, X-ray studies have been conducted on a beam-line X17A at the National Synchrotron Light Source at Brookhaven National Laboratory. Scattering intensity was measured from high energy X-rays at an energy of $67.419$ keV using the rapid acquisition pair distribution function (RAPDF) technique \cite{Chupas-JAC2003,Juhas-JAC2013}. A large area 2-D Perkin Elmer detector with $2048 \times 2048$ pixels and $200 \times 200$ $\mu$m pixel size was mounted orthogonal to the beam path with a sample-to-detector distance of $206.1371$ mm. Two samples have been measured for comparison: The sample of conventionally grown alongated micron-size crystallites of Mn$_{12}$ acetate and the sample of Mn$_{12}$ nanospheres described above. 

Structure factors and pair distribution functions for the two samples are shown in Fig. \ref{fig structure}. While distinct Bragg picks of Mn$_{12}$ acetate \cite{Langan-AC2001,Cornia-AC2002} are present in the micro-crystalline sample, a very diffused intensity has been found in the nanospheres. Same differences have been observed in the RAPDF, indicating high level of disorder in the nanospheres. They must be either extremely defected or amorphous. As we shall see from magnetic measurements, however, the Mn$_{12}$ molecules are well preserved in the nanospheres despite of the structural disorder. 

Amorphous or highly disordered structure of Mn$_{12}$ nanospheres raises the question whether the easy axes of Mn$_{12}$ molecules inside the spheres are aligned or disordered. One should notice in this connection that the direction of the easy axis is related to the orientation of the molecules while the Bragg peaks in the X-ray scattering are related to the translational order in a crystal. In an amorphous solid the orientation of local crystallographic axes is more robust than the translational order \cite{Nelson-JNCS1984}. It may spread well beyond the amorphous structure factor.  Evidence of the extended correlations in the orientation of local magnetic anisotropy axes has been previously reported in amorphous magnets \cite{CT-EPL1993}.  For the conclusions made in this paper it does not matter whether the easy axes of the molecules are disordered at the level of individual nanospheres, or due to the random orientation of nanospheres, or both. 

\section{Magnetic measurements}\label{magnetic}

Low-temperature magnetic measurements have been carried out on a compressed powdered sample inside a commercial rf-SQUID Quantum Design magnetometer. Before conducting the measurements, an attempt has been made to orient the nanoparticles by a high magnetic field at room temperature in the epoxy, as has been done previously with $\mu$m-size microcrystals. Contrary to the latter case, the orienting procedure applied to the nanospheres rendered no difference in the magnetization data. This can be understood from the fact that micron-size crystallites of Mn$_{12}$ acetate grow elongated along the easy magnetization axis. The resulting anisotropy of their magnetic susceptibility then enables the alignment of the crystallites by a high magnetic field in a Stycast, even though the molecules do not have magnetic moments at room temperature. This method, however, does not work for the Mn$_{12}$ nanoparticles of spherical shape. 

\begin{figure}[htbp!]
\includegraphics[width=8cm,angle=0]{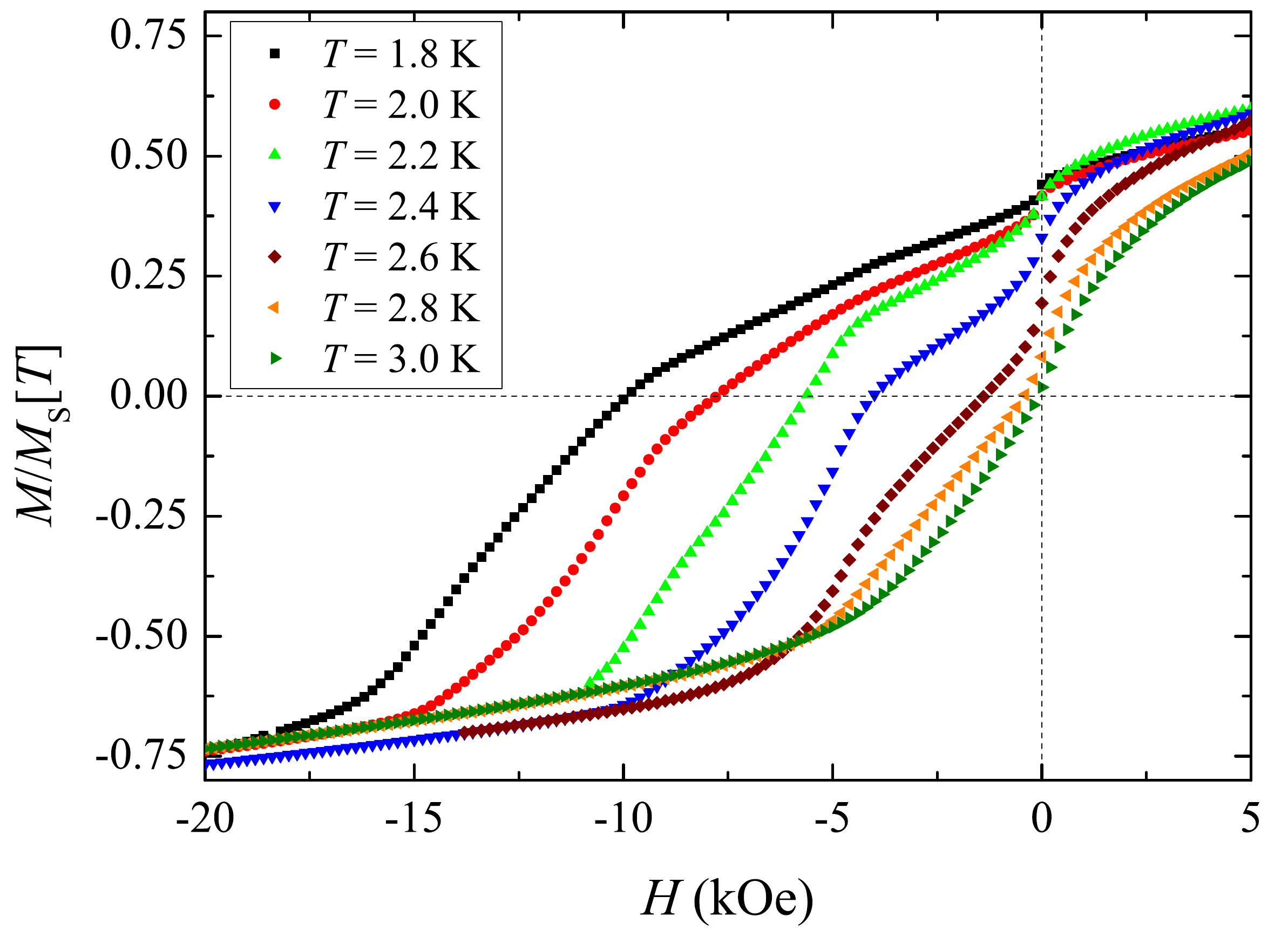}
\caption{Color online: $H < 0$ branch of the hysteresis curves of the disordered powder of Mn$_{12}$ nanospheres at different temperatures.}
\label{fig hysteresis}
\end{figure}
\begin{figure}[htbp!]
\includegraphics[width=8cm,angle=0]{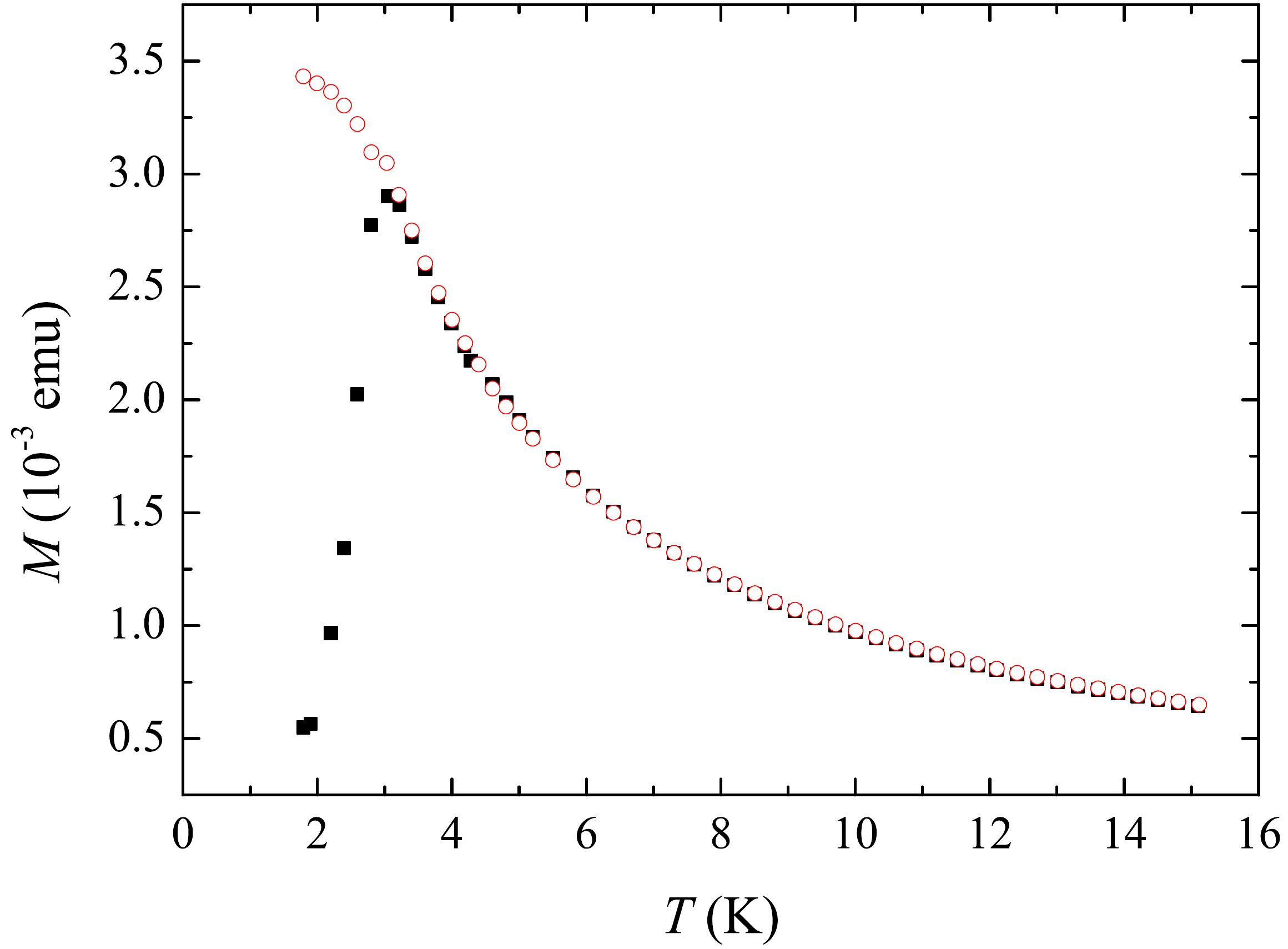}
\caption{Color online: Field-cooled (FC) and zero-field-cooled (ZFC) magnetization vs temperature curves for a disordered powder of Mn$_{12}$ nanospheres in the dc field of $100$ Oe.}
\label{fig FC-ZFC}
\end{figure}
\begin{figure}[htbp!]
\includegraphics[width=8cm,angle=0]{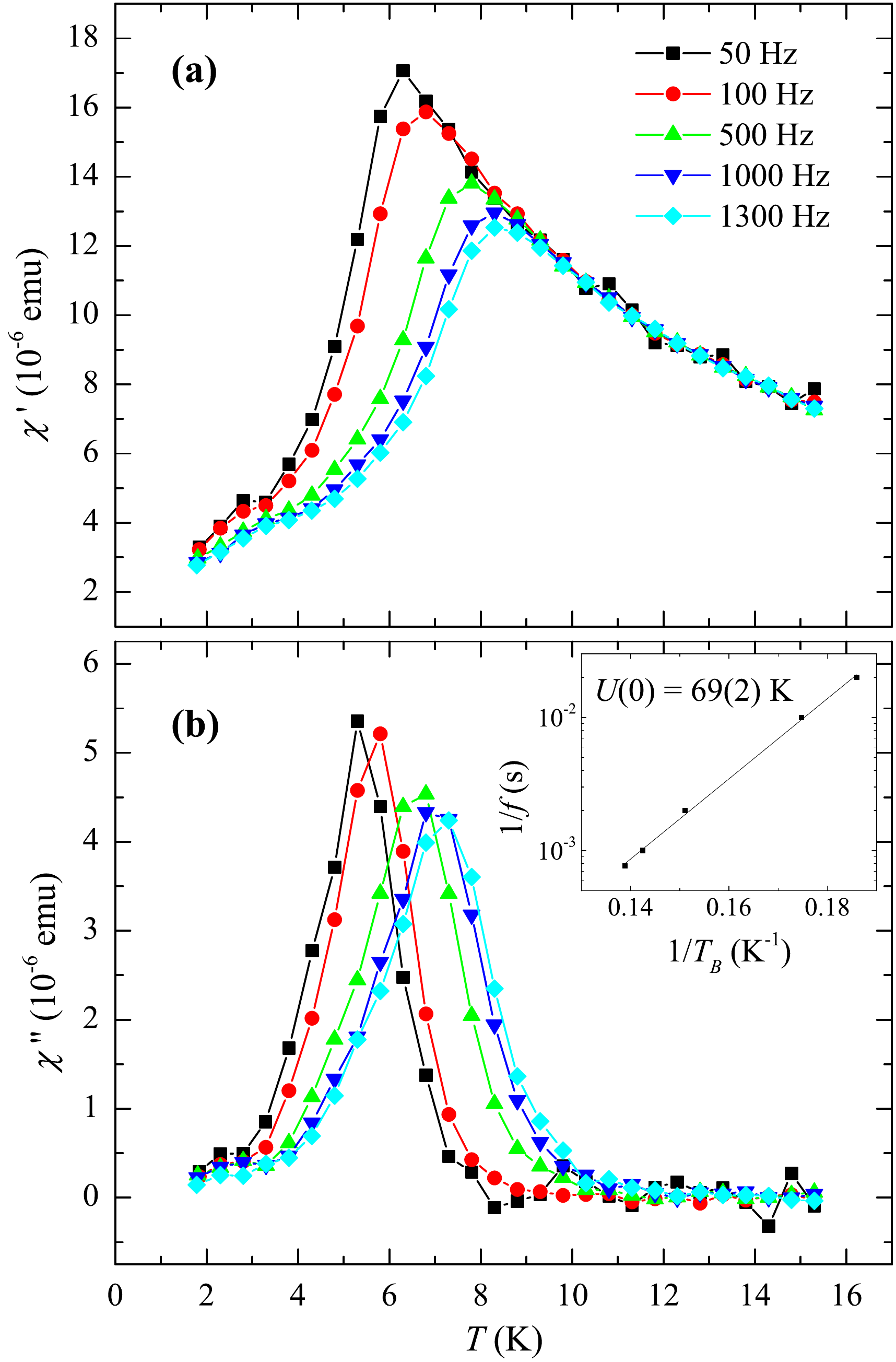}
\caption{Color online: Real part (a) and imaginary part (b) of the ac susceptibility in the frequency range $50$ Hz - $1.3$ kHz in the presence of a $3$ Oe ac field. The 
inset shows logarithmic dependence of the blocking temperature on $1/f$. The fitted value of the energy barrier in a zero field is $U(0)=69\pm2$ K.}
\label{fig AC susc}
\end{figure}

Negative branches of hysteresis curves are shown in Fig. \ref{fig hysteresis}. At the lowest temperature used, when thermal effects are weak, $M(0)$ is about one half of the saturation value, $M_S$. This corresponds to the magnetic moments of Mn$_{12}$ molecules looking randomly into a hemisphere, which follows from 
\begin{equation}
\frac{M(0)}{M_S} = \frac{1}{4\pi}\int_0^{2\pi}\textrm{d}\phi\int_0^{\pi/2}\sin\theta \textrm{d}\theta = \frac{1}{2}.
\end{equation}
The latter is true regardless of whether the disorder in the orientation of easy magnetization axes occurs at the level of individual Mn$_{12}$ molecules inside amorphous nanospheres or at the level of the nanospheres, each having a coherent orientation of the easy axes of the molecules. 

Field-cooled (FC) and zero-field-cooled (ZFC) magnetization curves are shown in Fig. \ref{fig FC-ZFC}. High-temperature behavior is a $1/T$ Curie law. 
Fig. \ref{fig AC susc} shows the ac susceptibility of the Mn$_{12}$ nanospheres. The maximum in both families of curves occurs at the blocking temperature, 
$T_B$, determined by the magnetic moments that reverse their orientation on the time scale of the measurement, $ t = t_0\exp[U(H)/T_B]$, with $t_0$ being the attempt time and $U(H)$ being the energy barrier. 
In the ac measurements $t \sim 1/f$, so that $T_B$ scales with the frequency of the ac field as $U(H)/\ln(1/t_0 f)$, which is in accordance with the data, see the inset in Fig. \ref{fig AC susc}b. The estimated values for the attempt time 
and the energy barrier at zero field are $t_{0}=(5\pm1)\times 10^{-8}$ s and $U(0)=69\pm2$ K, respectively.

One-maximum structure of the curves shown in Figs. \ref{fig FC-ZFC} and \ref{fig AC susc} contrasts with a pronounced two-maxima structure observed in Refs. \onlinecite{Imaz-ChemCom2008, Carbonera-ICA208}, with the second, lower maximum occurring at a lower temperature. The latter, according to the authors of Refs. \onlinecite{Imaz-ChemCom2008,Carbonera-ICA208}, where spherical nanoparticles of average size under $50$ nm were studied, corresponded to the presence of $40$\% fast-relaxing species of Mn$_{12}$ acetate. If such a species were present in our highly amorphous, larger spherical particles, it would have had a very low abundance. This simplifies the analysis of the data presented in the next section.

\begin{figure}[htbp!]
\includegraphics[width=8cm,angle=0]{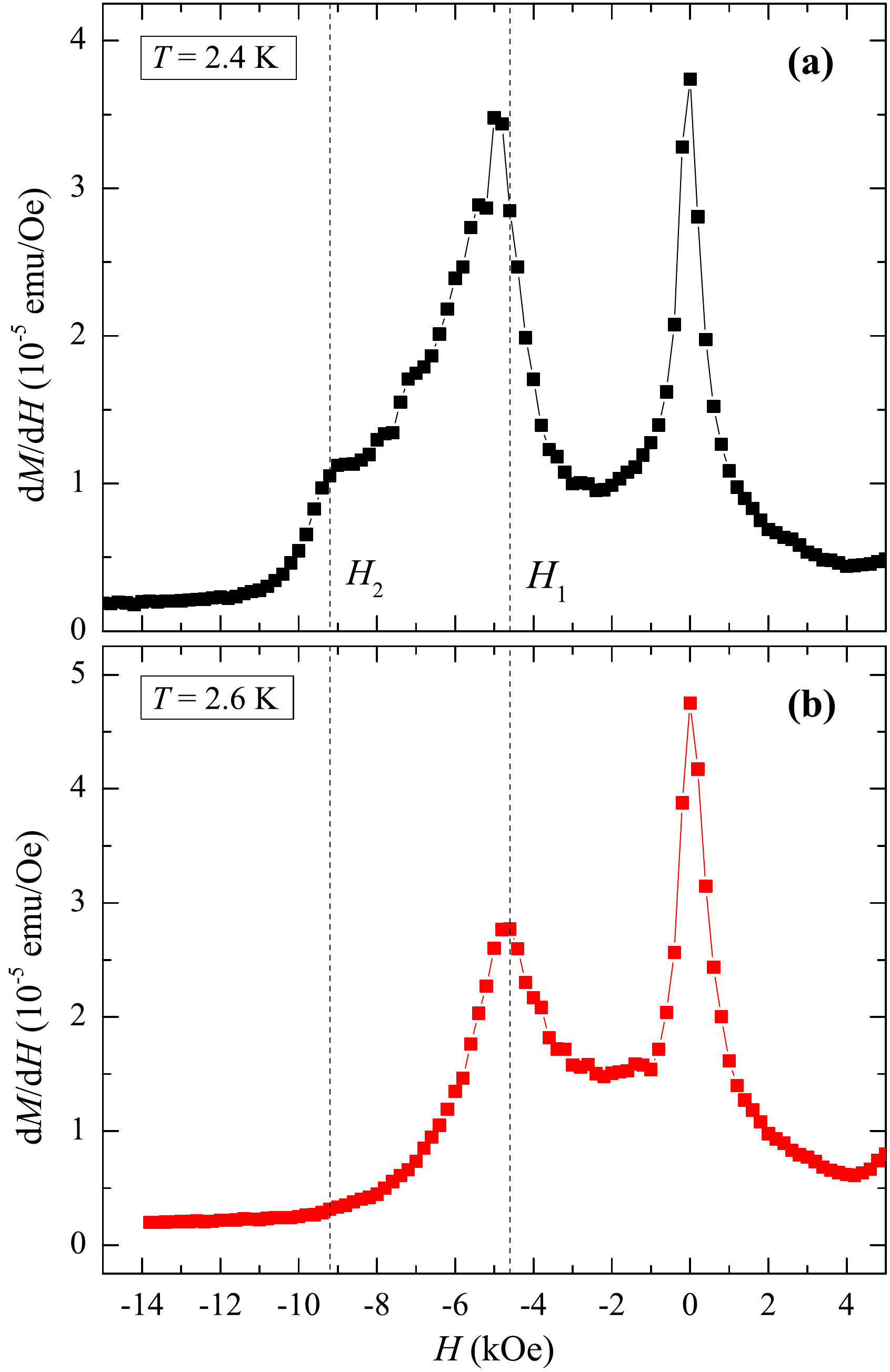}
\caption{Color online: d$M$/d$H$ for the $H < 0$ branch of the hysteresis curve of the disordered powder of Mn$_{12}$ acetate nanospheres: (a) $T = 2.4$ K, (b) $T = 2.6$ K.}
\label{fig dM/dH}
\end{figure}
Fig. \ref{fig dM/dH} shows the derivative, d$M$/d$H$, along the negative branch of the hysteresis curve shown in Fig. \ref{fig hysteresis}, for two temperatures. The observed zero-field maximum has the width of about $1$ kOe and is much more narrow than the zero-field maximum observed in Refs. \onlinecite{Imaz-ChemCom2008,Carbonera-ICA208}. There are clear maxima at $H_1 \approx 4.6$ kOe and $H_2 \approx 9.2$ kOe as in a Mn$_{12}$ acetate crystal.

\section{Analysis of the magnetization data}\label{analysis}

At first glance, the presence of the maxima in the d$M$/d$H$ plot, Fig. \ref{fig dM/dH}, that correlate with the typical tunneling maxima of Mn$_{12}$ acetate at $H_n = nH_1$, with $H_1 \approx 4.6$ kOe,  is surprising given the random orientation of the easy magnetization axes of the molecules. Simple analysis presented below shows however that the tunneling maxima can and should be observed in d$M$/d$H$ obtained from a sample of randomly oriented magnetic molecules. Let us look closer at the magnetization curve in Fig. \ref{fig hysteresis} that corresponds to the sweep from a positive to a negative field. 

\begin{figure}[htbp!]
\includegraphics[width=8.8cm,angle=0]{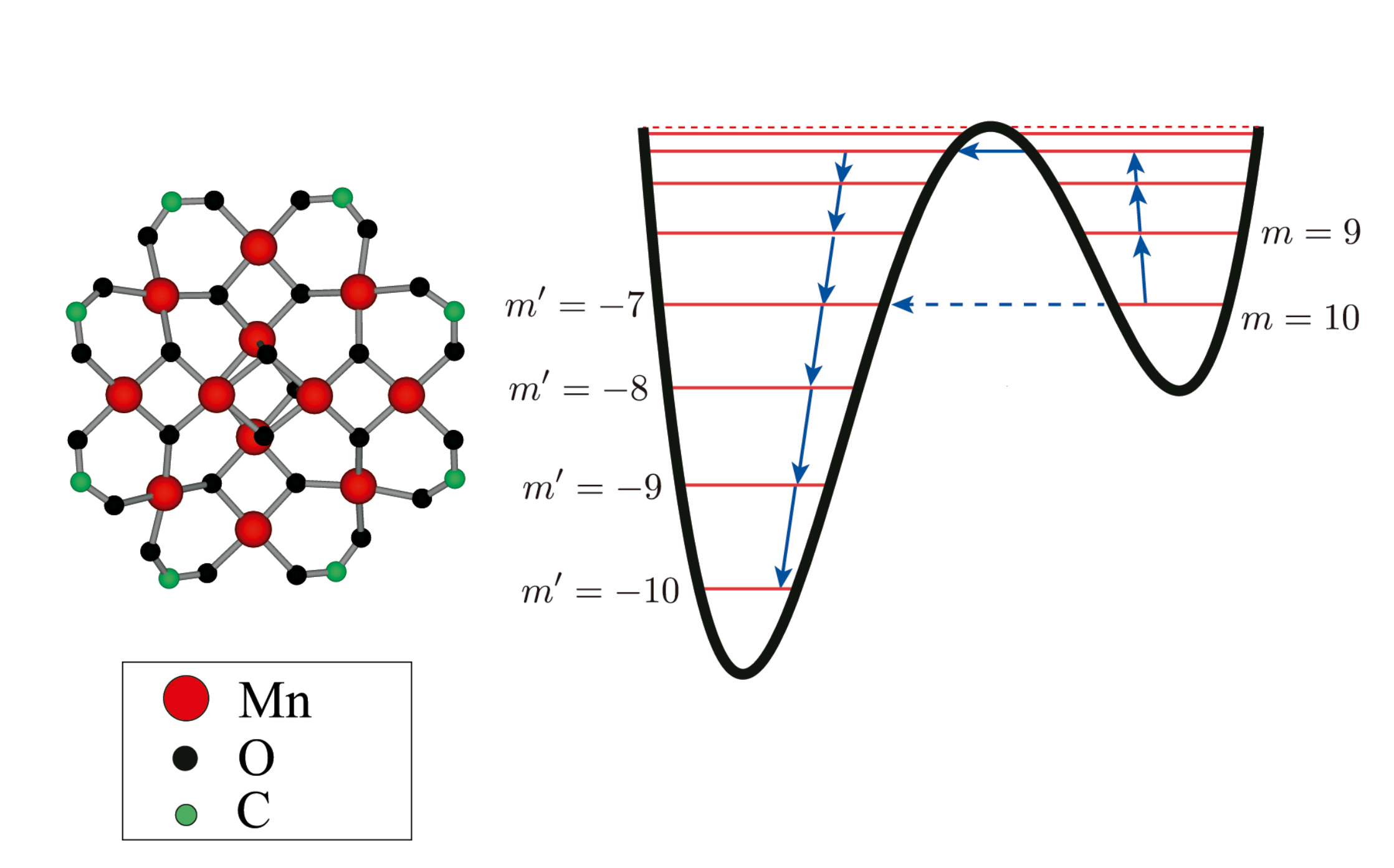}
\caption{Thermally assisted tunneling between resonant spin levels in a Mn$_{12}$ molecule.}
\label{fig resonant-tunneling}
\end{figure}
Thermally assisted spin tunneling illustrated by Fig. \ref{fig resonant-tunneling} takes place as the field goes into the negative territory, $H < 0$. If the easy axis of a particular molecule makes the angle $\theta$ with the negative direction of the field, then the condition of the $n$-th resonance is
\begin{equation}
H = -\frac{H_n}{\cos \theta}, \quad 0 \leq \theta \leq \frac{\pi}{2}
\end{equation}
Thus, at any value of the field $H$ the molecules that are on the $n$-th resonance satisfy
\begin{equation}
\cos \theta = -\frac{H_n}{H}
\end{equation}

The zero-field maximum in d$M$/d$H$ that has been observed in all cases has an obvious explanation. At $H = 0$ all molecules are close to the $n = 0$ resonance up to the detuning effect of local dipolar and hyperfine fields that provide the finite width of the zero-field resonance. The latter is in the ballpark of $1$ kOe or below. Here we  focus on the subsequent resonances $n = 1,2,...$ that have not been reported for disordered powders in the past. 

Since the element of the spherical volume is proportional to $\sin\theta \textrm{d}\theta = -\textrm{d}\cos\theta$, the total number of molecules that enter the resonance condition when the field increases in the negative direction by a small increment d$H < 0$ is 
\begin{equation}
\textrm{d}N_{\textrm{res}} \propto -\textrm{d}\cos\theta \propto -\frac{\textrm{d}H}{H^2}\sum_{H_n \leq |H|} H_n > 0
\end{equation}
At low temperature, when the field sweep is sufficiently slow, the change in the magnetization, d$M$, should be proportional to $-\textrm{d}N_{\textrm{res}}$, 
\begin{equation}
\textrm{d}M \propto -\textrm{d}N_{\textrm{res}} \propto \frac{\textrm{d}H}{H^2}\sum_{H_n \leq |H|} H_n < 0.
\end{equation}
This gives
\begin{equation}
\frac{\textrm{d}M}{\textrm{d}H} \propto \frac{1}{H^2}\sum_{H_n \leq |H|} H_n > 0
\end{equation}

The function in the right-hand-side of this equation has a jump-wise behavior. At $H_1 \leq |H| < H_2 = 2H_1$ it behaves as $H_1/H^2$. At $H_2 \leq |H| < H_3 = 3H_1$ it behaves as $3H_1/H^2$. At $H_3 \leq |H| < H_4 = 4H_1$ it behaves as $6H_1/H^2$, and so on. Its value at $H = H_n$ equals 
\begin{equation}
\frac{1}{H_n^2}\sum_{H_n \leq |H|} H_n  = \frac{1}{n^2 H_1}\sum_{k=1}^{n} k = \frac{1}{2H_1}\left(1 + \frac{1}{n}\right).
\end{equation}

\begin{figure}[htbp!]
\includegraphics[width=8cm,angle=0]{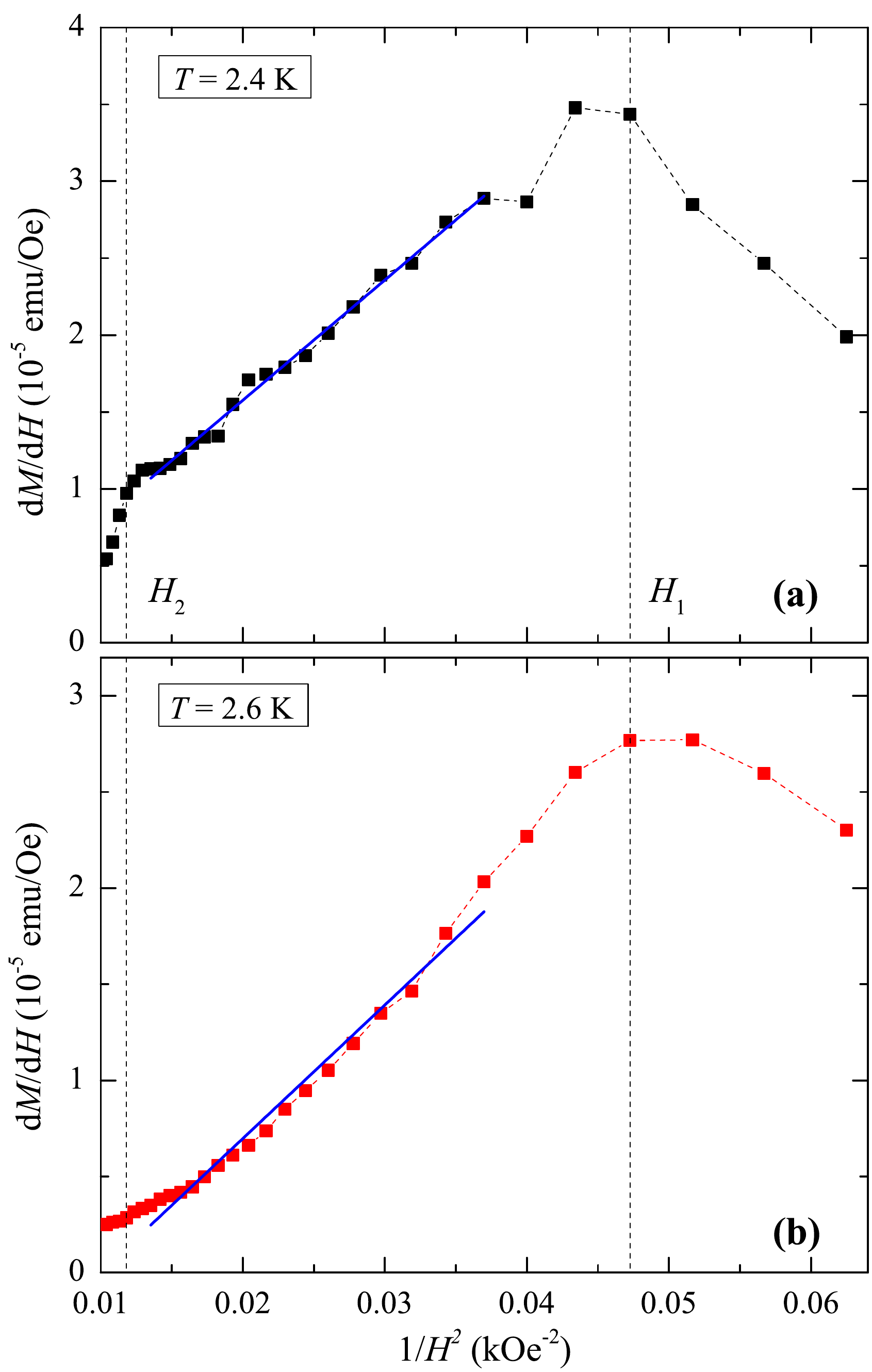}
\caption{Color online: The $1/H^2$ fit (blue line) of d$M$/d$H$ between first and second spin tunneling resonances for a disordered powder of Mn$_{12}$ nanospheres: (a) $T = 2.4$ K, (b) $T = 2.6$ K.}
\label{fig 1/H^2}
\end{figure}
This analysis leads to the following conclusion. Spin tunneling resonances in a system of randomly oriented magnetic molecules can be seen as maxima in d$M$/d$H$. However, only the $H = 0$ maximum is a narrow tunneling resonance for all molecules within dipolar/hyperfine window. Each subsequent $n$-th ``resonance'' begins at $H = -H_n$ and extends to all $H < -H_n < 0$ due to different resonance conditions for differently oriented molecules. If one neglects broadening due to dipolar and hyperfine fields, these ``resonances'' begin with a vertical jump in d$M$/d$H$ at $H = - H_n$, followed by the $1/H^2$ decrease of d$M$/d$H$. Dipolar and hyperfine fields must smear the initial vertical rise of the ``resonance'' by up to $1$ kOe. However, the $1/H^2$ post-resonance ($H < -H_n$) decrease of d$M$/d$H$ over the $4.6$ kOe range from $H = -H_n$ to $H = -H_{n+1}$ must be apparent. This is seen in Fig. \ref{fig 1/H^2} that shows the fit of the post-resonance d$M$/d$H$ by $1/H^2$ after crossing the first resonance. In the kelvin temperature range the effect becomes less pronounced at higher $n$ due to the increased thermal relaxation on lowering the barrier by the field .

\begin{figure}[htbp!]
\includegraphics[width=8cm,angle=0]{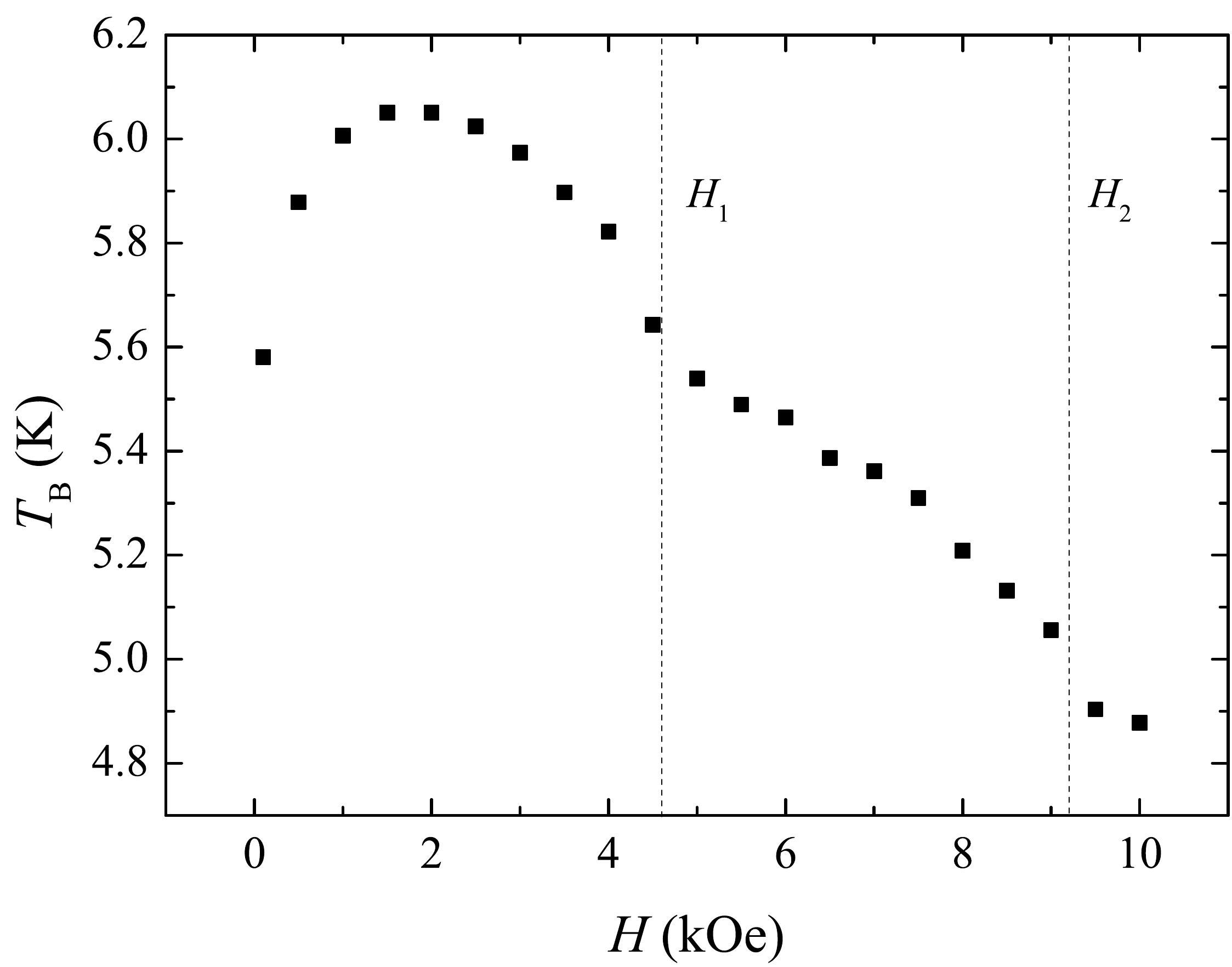}
\caption{Color online: Dependence of the blocking temperature on the dc field in the ac field of $100$ Hz.}
\label{fig blocking-H}
\end{figure}
Note that for the same reason as discussed above the evidence of spin tunneling in a disordered system can also be obtained by plotting the dependence of the blocking temperature on the dc magnetic field, see Fig. \ref{fig blocking-H}. The blocking temperature, $T_B$, is proportional to the energy barrier for the spin reversal, $U(H)$. At $H = 0$ the barrier is lowered due to thermally assisted quantum tunneling between resonant spin levels near the top of the barrier, see Fig. \ref{fig resonant-tunneling}. The number of molecules on the zero-field resonance at a particular value of the field progressively drops as the field approaches the boundary of the dipolar/hyperfine window, which for Mn$_{12}$ has a width of the order of $1$ kOe. At that boundary $T_B$ is approximately determined by the full $69$ K classical energy barrier for the superparamagnetic spin flip. As the field continues to increase, the field dependence of the blocking temperature is dominated by the classical reduction of the energy barrier due to the growing Zeeman energy, which leads to the decrease of $T_B$ on $H$. When the field reaches $H_1 \approx 4.6$ kOe, randomly oriented Mn$_{12}$ molecules begin to enter the first resonance. The evidence of that is clearly seen in Fig. \ref{fig blocking-H} in the change of the derivative of $T_B$ on $H$ at $H = H_1$. 

\section{Discussion}\label{discussion}

We have studied spin tunneling effects in amorphous nanospheres of Mn$_{12}$ acetate. Samples have been characterized by chemical, infrared, TEM, X-ray, and magnetic methods. While the structure of Mn$_{12}$ molecules appears to be the same as in a crystalline sample, their easy magnetization axes are completely disordered. Random orientation of the easy axes has been confirmed by magnetic measurements. It occurs either at the level of individual molecules inside the nanospheres, or at the level of randomly oriented nanospheres, or both. Isotropic magnetic susceptibility of the nanospheres at room temperature, when Mn$_{12}$ molecules do not possess magnetic moments, excludes any possibility of aligning their easy axes in a high magnetic field as was done for the sample of Mn$_{12}$ acetate consisting of elongated microcrystals \cite{Friedman-PRL1996}. 

Measurements of the FC and ZFC magnetization curves, and of the ac susceptibility, revealed presence of only one magnetic species of Mn$_{12}$ acetate, which is in contrast with previous measurements on nanospheres containing comparable amounts of slow and fast relaxing species \cite{Imaz-ChemCom2008,Carbonera-ICA208}. The striking feature of the magnetization curve is the presence of well-defined sharp tunneling maxima in d$M$/d$H$ in a sample with random orientation of easy magnetization axes of the molecules. The maxima occur at the conventional resonant fields, $H_n = -nH_1$, with $H_1 \approx 4.6$ kOe. 

\begin{figure}[htbp!]
\includegraphics[width=8cm,angle=0]{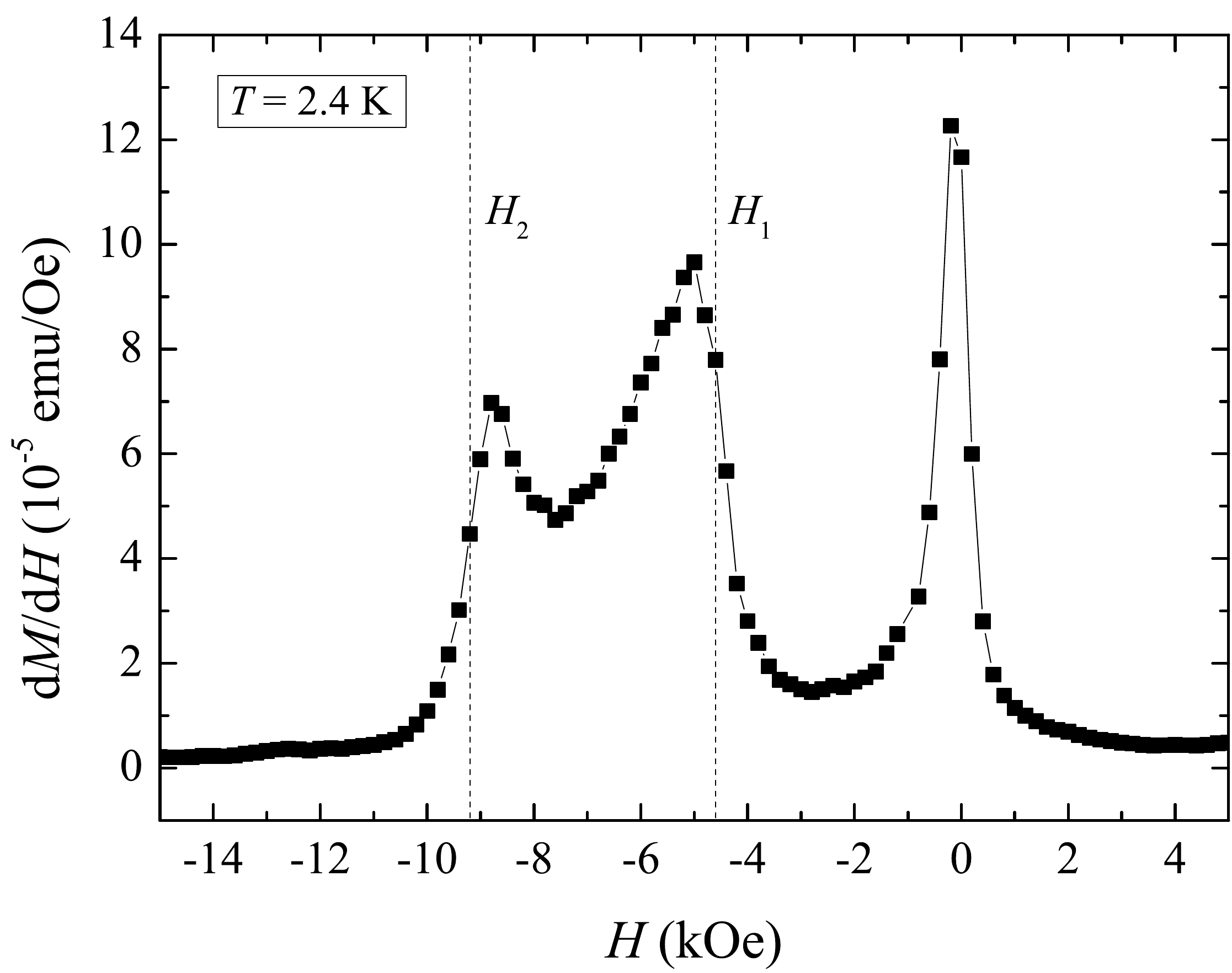}
\caption{Color online: Field derivative of the magnetization in a sample of randomly oriented micron-size crystallites of Mn$_{12}$ acetate.}
\label{fig microcrystals}
\end{figure}
Quantitative explanation to this observation is provided. At any value of the field $H$ the molecules that are on the $n$-th resonance have their easy axis at an angle $\theta$ with the field, satisfying $\cos \theta = -H_n/H$. Thus, as soon as the field enters the range $H \leq - H_n$ a new set of molecules that satisfy the $n$-th resonance condition begins to reverse their magnetization.  This results in the change of the slope of $M(H)$ every time when the field reaches the value $H_n = -nH_1$, and in the corresponding peaks in d$M$/d$H$. However, only the $n=0$ peak is the same tunneling resonance as in a crystalline sample, when all molecules achieve the resonance condition inside the dipolar/hyperfine field window. As our analysis show, all subsequent resonances are characterized by a steep jump of d$M$/d$H$ at $H_n = -nH_1$ followed by a $1/H^2$ decrease. This conclusion agrees with the experimental data.

Inevitably, our findings raise the following question. Is the effort to grow a single crystal or the laborious procedure of aligning microcrystals, as was done in Ref. \onlinecite{Friedman-PRL1996} that reported the discovery of the resonant spin tunneling in Mn$_{12}$ acetate, really necessary to obtain the evidence of spin tunneling? As has been illustrated by this paper, the answer is apparently no. To further illustrate this point we conducted measurements of the non-oriented powder of micron-size crystals identical to the powder used in Ref. \onlinecite{Friedman-PRL1996} where microcrystals were oriented in an epoxy prior to taking magnetic measurements. The d$M$/d$H$ plot obtained from the powder sample without any orientation effort is presented in Fig. \ref{fig microcrystals}. It clearly shows three tunneling maxima. 

We, therefore, conclude that while single crystals are good for the in depth study of spin tunneling, the robust evidence of the effect can be obtained from the magnetization curve of any micro-crystalline sample as long as the magnetic molecules preserve their structure. This should greatly simplify the effort by chemists and physicists to track spin tunneling effects in newly synthesized molecular magnets. 

\section{Acknowledgements}

The work at the University of Barcelona has been supported by the Spanish Government Projects No. MAT2008-04535 and MAT2012-30994. 
S.L. acknowledges financial support from the FPU Program of Ministerio de Educaci\'{o}n, Cultura y Deporte of the Spanish Government. I.I. and J.E. thank the MINECO for the Ram\'{o}n y Cajal contract and the FPI fellowship, respectively. 
The work of E.M.C. at Lehman College is supported by the U.S. National Science Foundation through grant No. DMR-1161571.

\end{document}